\documentstyle[pre,aps,epsf]{revtex}\draft
\begin{document}

%\twocolumn[\hsize\textwidth\columnwidth\hsize\csname@twocolumnfalse\endcsname
\title{High-Acceleration Patterns in Thin Vibrated Granular Layers} 
\author{D. Blair,  I.S.  Aranson, G.W. Crabtree and V. Vinokur}
\address{Argonne National Laboratory,
9700 South  Cass Avenue, Argonne, IL 60439}
\author{L.S. Tsimring}
\address{
Institute for Nonlinear Science, University of California, San Diego, La
Jolla,
CA 92093-0402}
\author{C. Josserand} 
\address{James Frank Institute, University of Chicago,
5640 S. Ellis Av., Chicago, IL60637} 

\date{\today}
\maketitle
\begin{abstract}
Theoretical and experimental study of high-acceleration
patterns in vibrated granular layers is presented.  The order parameter 
model based
on parametric Ginzburg-Landau equation is used to describe strongly
nonlinear excitations including hexagons, interface between flat
anti-phase domains and new localized objects, {\em super-oscillons}.
The experiments confirmed the existence of super-oscillons and bound
states of super-oscillons and interfaces. On the basis of 
order parameter model we predict analytically and confirm 
experimentally that the additional subharmonic 
driving results in controlled motion of the interfaces.
\end{abstract}
%\narrowtext
%\vskip1pc]

The collective dynamics of granular materials is a subject of current
interest \cite{jnb,swin1,swin2,swin3,inter}. The intrinsic dissipative
nature of the interactions between the constituent macroscopic particles
give rise to several basic properties specific to granular substances
and setting granular matter apart from the conventional gaseous, liquid,
or solid states.

Driven granular systems  manifest collective fluid-like behavior:
convection, surface waves, and pattern formation (see e.g.  \cite{jnb}).
One of the most fascinating  examples of  this collective dynamics is
the appearance of long-range coherent patterns and localized excitations
in vertically-vibrated thin granular layers
\cite{swin1,swin2,swin3,inter,jag3,mm}.  The particular pattern is
determined by the interplay between driving frequency $f$ and
acceleration of the container $\Gamma= 4 \pi^2 {\cal A }  f^2/g$ (${\cal
A} $ is the amplitude of oscillations, $g$ is the gravity acceleration)
\cite{swin1,swin2}.

Patterns appear  at   $\Gamma\approx 2.4$  almost independently of the
driving frequency $f$.  At small frequencies $f< f^* $
\cite{swin2,swin3}  the transition is subcritical (hysteretic), leading
to formation of squares.  In the hysteretic region, localized
excitations  such as individual {\it oscillons} and various bound states
of oscillons appear as $\Gamma$ is decreased.  For  higher frequencies
$f >f^* $ the onset pattern is stripes, and at the frequency slightly
higher than $f^*$ the transition becomes supercritical. Both squares and
stripes, as well as oscillons, oscillate at half of the driving
frequency $f/2$. At higher acceleration ($\Gamma>4 $), stripes and
squares become unstable, and hexagons appear instead. Further increase
of acceleration at $\Gamma \approx 4.5$ converts hexagons into a
domain-like structure of flat layers oscillating with frequency $f/2$
with opposite phases. Depending on parameters, interfaces separating
flat domains, are either smooth or ''decorated'' by periodic
undulations.  For $\Gamma >5.7$  various quarter-harmonic patters
emerge.

The pattern formation in thin layers of granular material was studied
theoretically by several groups. Direct molecular dynamics simulations
\cite{bizon,bizon1} (see also \cite{luding}) reproduced a majority of patterns
observed in experiments and many features of the bifurcation diagram,
although until now have not yielded oscillons and interfaces.
Hydrodynamic and phenomenological models \cite{theor,crawford} reproduced
certain  experimental features, however neither of them offered a
systematic description of the whole rich variety of the observed
phenomena.  In Refs. \cite{at,atv} we introduced the order parameter
characterizing the complex amplitude of sub-harmonic oscillations.  The
equations of motion following from the symmetry arguments and mass
conservation reproduced essential  phenomenology of patterns near the
threshold of primary bifurcation:  stripes, squares, and,  
localized objects,   oscillons.

In this paper we describe high acceleration patterns on the basis of the
same order parameter model and compare it with experimental
observations.  Our preliminary results were published earlier in  Refs.
\cite{inter,atv,blair}. Here we show that at large amplitude of driving both
hexagons and interfaces emerge. We find morphological instability
leading to the formation of ``decorated''  interfaces. We study the
motion of the interface under the influence of a small subharmonic
component in the driving acceleration.  We also find a new 
localized structure, "super-oscillon", which exists for high
acceleration values. We discuss possible mechanisms of saturation of
the interface instability. Our experimental results demonstrate
the  existence of super-oscillons and bound states of
super-oscillons and interfaces. They also confirm theoretical
predictions of the externally controlled interface motion.

The structure of the article is following. In Sec. I we introduce our
phenomenological order-parameter model. In Sec. II  we develop the 
weakly-nonlinear theory  for the flat period-doubled state of the
vibrated layer. In Sec. III we analyze the interface solution and study
its stability with respect to transverse perturbations.  In Sec. IV we
discuss new localized solutions. In Sec. V we present the combined
theoretical phase diagram of the model.  In Sec. VI we demonstrate that
in a certain limit our model can be reduced to the real Swift-Hohenberg
equation.  This equation also possesses a similar variety of patterns
including stripes, hexagons, stable and unstable interface solutions, as
well as localized solutions.  
In Sec. VII we demonstrate that additional subharmonic driving results in 
drift of the interface with the velocity determined by the amplitude of 
the driving and the direction determined by the relative phase. 
In Sec. VIII we present our experimental
results. It includes the study  phase diagram  and the 
effect of additional subharmonic 
driving. 
Section IX summarizes our conclusions.

\section{Parametric Ginzburg-Landau Equation} 

The essence of the  model \cite{at,atv} is  the order parameter equation
for the complex amplitude $\psi$ of parametric layer oscillations
$h=\psi\exp(i\pi f t)+c.c.$ at the frequency
$f/2$  coupled with the equation for the thickness of the
layer $\rho$ averaged over the period of vibrations, 
\begin{eqnarray}
\partial_t\psi&=&\gamma\psi^*-(1-i\omega)\psi+(1+ib)\nabla^2\psi
-|\psi|^2\psi-\rho\psi \label{eq0a}\\
\partial_t\rho&=&  \zeta \nabla\cdot(\rho\nabla|\psi|^2)+\beta\nabla^2\rho
\label{eq0b}
\end{eqnarray}
Here $\gamma$ is the normalized amplitude of forcing at the resonant
frequency $f$. 
Linear terms in  Eq. (\ref{eq0a})  can be obtained
from the complex growth rate for infinitesimal periodic layer
perturbations $h \sim  \exp[ \Lambda(k) t + i k x]$. Expanding
$\Lambda(k)$ for small $k$, and keeping only two leading terms in the
expansion $\Lambda (k) = -\Lambda_0   -\Lambda_1 k^2$ we obtain  the
linear terms in Eq. (\ref{eq0a}), where $b = Im \Lambda _1/ Re \Lambda_1$
characterizes ratio of dispersion to diffusion
and parameter 
$\omega = (\Omega_0- \pi f ) /Re \Lambda _0 $, $\Omega_0 = -
Im \Lambda_0$ characterizes the frequency of the driving.
In Eq. (\ref{eq0b}) 
$\zeta, \beta$ are transport coefficients for $\rho$.
Slowly-varying thickness of the layer $\rho$ controls the dissipation
rate (the last term in Eq.(\ref{eq0a})). The second equation
(\ref{eq0b}) describes re-distribution of the averaged thickness due to the
diffusive flux $\propto -\nabla \rho$, and an additional flux $\propto
-\rho\nabla |\psi|^2$ caused by the spatially nonuniform vibrations of
the granular material. This coupled model was used in Refs.\cite{at,atv}
to describe the pattern selection near the threshold of the primary
bifurcation.  It was shown that at small $\zeta \rho_0\beta^{-1}$ (which
corresponds to low frequencies and thick layers), the primary
bifurcation is subcritical and leads to the emergence of square
patterns. For higher frequencies and/or thinner layers, transition is
supercritical and leads to roll patterns. At intermediate frequencies
($f\approx f^*$), the stable localized solutions of
Eqs.(\ref{eq0a}),(\ref{eq0b}) corresponding to isolated {\em oscillons} 
and variety of bound states were
found in agreement with experiment. 

In this paper we focus on high-acceleration
patterns at high frequencies.
In Ref. \cite{at} it was indicated  that the density transport coefficient 
$\beta$ is proportional to the energy of plate vibration ($\propto f^2$), and, 
therefore,  it should  increase  with the driving frequency $f$.
As a result, for high frequencies the coupling
between  $\rho$ and  $\psi$ becomes less relevant, and one can assume 
$\rho=$const and exclude it from Eq.(\ref{eq0a}) by rescaling. Then  
the model can be reduced to a single order-parameter equation
\begin{eqnarray}
\partial_t\psi&=&\gamma\psi^*-(1-i\omega)\psi+(1+ib)\nabla^2\psi
-|\psi|^2\psi
\label{eq1}
\end{eqnarray}
Eq.(\ref{eq1}) also describes the evolution of the order parameter for the
parametric instability in vertically oscillating fluid layers  (see
\cite{vinals,kiyashko}).  
%Linear terms in this equation can be obtained
%from the complex growth rate for infinitesimal periodic layer
%perturbations $h \sim  \exp[ \Lambda(k) t + i k x]$. Expanding
%$\Lambda(k)$ for small $k$, and keeping only two leading terms in the
%expansion $\Lambda (k) = -\Lambda_0   -\Lambda_1 k^2$ we obtain  the
%linear terms in Eq. (\ref{eq1}), where $b = Im \Lambda _1/ Re \Lambda_1$
%and $\omega = (\Omega_0- \pi f ) /Re \Lambda _0 $, where $\Omega_0 = -
%Im \Lambda_0$. 

It is convenient to shift  the phase of the complex order
parameter via $\tilde{\psi}=\psi\exp(i\phi )$ with $\sin
2\phi=\omega/\gamma$.  The equations  for real and imaginary part
$\tilde \psi = A + i B$ are:
\begin{eqnarray}
\partial_t A &=&(s -1) A -  2  \omega B - ( A^2+ B^2) A +\nabla^2 ( A- b B)
%-\rho A
\label{eq2a} \\
\partial_t B  &=& -  (s + 1)  B - ( A^2+ B^2) B +\nabla^2 ( B + b  A )
%-\rho B
\label{eq2b}
\end{eqnarray}
where $s^2 = \gamma^2 -\omega^2$. 

\section{Stability of uniform states} 

At $s<1$, Eqs.
(\ref{eq2a}),(\ref{eq2b}) has only one trivial uniform state $A=0,\
B=0$, At $s>1$, two new uniform states  appear, $A=\pm \sqrt{s-1}, B=0$.
The onset of these states corresponds to the period
doubling of the layer flights sequence,  observed in experiments
\cite{swin1} and predicted by the simple  inelastic ball model
\cite{swin1,metha}. Signs $\pm$ reflect two relative phases of layer
flights with respect to container  vibrations\cite{note4}.

The trivial flat state $A=B=0$ loses stability if the following condition fulfilled
(compare \cite{at}): 
\begin{equation} 
\gamma^2> \frac{(\omega+b)^2 }{1+b^2} 
\label{cond1} 
\end{equation} 
Because of the symmetry of Eq.(\ref{eq1}), small perturbations from the trivial 
state lead to the formation of a periodic  sequence of rolls or stripes. 

Let us  analyze the stability of the non-trivial  states $A=\pm \sqrt{s-1},
B=0$ 
with respect to small perturbations with wavenumber $k$ 
%(for definetness 
%we will focus on a state with sign $+$)
\begin{equation}
{A \choose  B}  = {\pm \sqrt{s-1} \choose  0 }  + {U_k \choose  V_k }
\exp [\lambda(k) t +  i k x ].
\label{eq4}
\end{equation}
At the threshold the uniform state loses its stability
with respect to periodic modulations with the critical wavenumber $k_c$
at $s < s_c$ (correspondingly, $\gamma< \gamma_c$), where
\begin{eqnarray}
s_c &=&\frac{ \sqrt{(1+\omega^2)(1+b^2)}-\omega+b}{2b}
\label{sc}  \\
k_c^2 &=&-  \frac{2s-1-\omega b} { 1+b^2}
\label{kc}
\end{eqnarray}

Small perturbations with every
direction of the wavevector grow at the same rate. The resultant
selected pattern is determined by the nonlinear competition between
the modes. In the presence of the reflection symmetry $\psi\to
-\psi$, quadratic nonlinearity is absent, and cubic nonlinearity near
the trivial state favors stripes corresponding to a single mode.
Near the fixed points $A=\pm \sqrt{s-1} , B=0$ the reflection symmetry for
perturbations $U \to-U,\ V \to -V$ is broken, and
hexagons emerge  at the threshold of the instability.
To clarify  this point
we perform weakly-nonlinear analysis  of Eqs. (\ref{eq2a}),(\ref{eq2b})
for  $s = s_c - \epsilon$, and $\epsilon \ll 1 $.
At $\epsilon \to 0$, the variables $U$ and $V$ are related as
in linear system:
\begin{eqnarray}
{U \choose  V} & =&  { 1 \choose \eta}  \Psi \nonumber \\
\Psi & = & \sum A_j \exp [ i {\bf k r } ] + c.c.,  \label{eq11}
\end{eqnarray}
where $| k| = k_c$ and $
\eta  =  \left[ 2(s_c -1) + k^2_c\right]/( b k^2_c-2\omega)$.

Corresponding  adjoint eigenvector is
\begin{eqnarray}
 {U^+ \choose  V^+ }  =  { 1 \choose \eta ^+  }  
 \label{eq13}
\end{eqnarray}
where 
$\eta^+   =  - \left[ 2(s_c -1) + k^2_c\right]/ b k^2_c$.
Substituting Eq. (\ref{eq11}) into Eqs. (\ref{eq2a}),(\ref{eq2b})  and
performing the
orthogonalization, we obtain  equations for the slowly-varying
complex amplitudes
$A_j$, $j=1,2,3$ (we assume only three waves with triangular symmetry, favored
by quadratic nonlinearity)
\begin{eqnarray}
\partial_t  A_j &=&  2 \epsilon A_j + a_2 A_{j+1}^* A_{j-1} ^*   \nonumber \\
&-&a_3 \left ( | A_j|^2+ 2( |A_{j-1}| ^2+ |A_{j+1}|^2 ) \right ) A_j,
\label{eq14}
\end{eqnarray}
where the coefficients $a_2,a_3$ are
\begin{eqnarray}
a_2  =  \pm 2 \sqrt{s-1} \left(2 + \frac{1+\eta^2}{1+ \eta \eta^+ } \right) \;,\;
a_3  =  3 ( 1+ \eta \eta^+).
\label{coef}
\end{eqnarray}

Eqs. (\ref{eq14}) are well-studied (see \cite{ch}). There are three critical
values of
$\epsilon$: $ \epsilon_A=-a_2^2/40a_3,\ \epsilon_R=a_2^2/2a_3,$ and
$\epsilon_B=2a_2^2/a_3$.
The hexagons are stable for $\epsilon_A < \epsilon < \epsilon_B$, and
the stripes are stable for $\epsilon > \epsilon_R$.
Thus, near $s=s_c$ the model exhibit stable  hexagons \cite{belg}.
Since we have two symmetric fixed points,
both up- and down-hexagons co-exist. For smaller $s$
 stripes are stable, and for larger $s$, flat
layers are stable, in agreement with observations\cite{swin2,blair}, as
well as direct numerical simulations of Eq.(\ref{eq1})\cite{numerics}.
The above  analysis requires the values of $\epsilon_{A,B,R}$ to  be
small. For parameters $\omega, b= O(1)$, this requirement  is satisfied
for $\epsilon_A  $, but not for  $\epsilon_{B,R}$. The estimates can be
improved by substituting   $s=s_c+\epsilon$ instead of
$s=s_c$  in (\ref{coef}). The resulting range of stable hexagons is
plotted in the phase diagram $(\omega,\gamma)$ (see below Fig.\ref{phase}). 

\section{Interface Solution} 

\begin{figure}
\centerline{\epsfxsize= 8.2cm\epsfbox{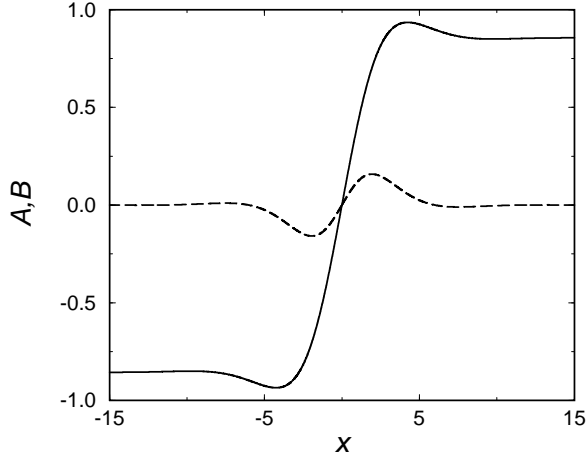}}
\caption{ Interface solution to Eqs. \protect{\ref{int_s}} 
obtained numerically for $\omega=1, b=4, \gamma=2$. Solid line corresponds 
to $A$ and dashed line to $B$, respectively.} 
\label{int_num} 
\end{figure}

At $s>1$,  Eqs. (\ref{eq2a}),(\ref{eq2b}) have an  interface solution
connecting two uniform states
$A=\pm \sqrt{s-1},\ B=0$. Let $x$ be the coordinate perpendicular to 
the interface  and $y$ is parallel to the interface. 
The interface is the stationary solution to Eqs. (\ref{eq2a}),(\ref{eq2b}):
\begin{eqnarray}
(s -1) A_0 -  2  \omega B_0 - ( A^2_0+ B^2_0) A_0 +A_{0xx} - b B_{0xx}&=&0,
\nonumber \\
 -  (s + 1)  B_0 - ( A^2_0+ B^2_0) B_0 +B_{0xx} + b  A_{0xx}&=&0.
\label{int_s}
\end{eqnarray}
For $b=0$ Eqs. (\ref{int_s}) have a  solution  of the form $  A_0= \pm
\sqrt{s-1} \tanh (\sqrt{s-1} x /2 ), B_0=0$. For $b\ne 0$ the solution is
available only numerically.  We used shooting matching technique to find
this solution in the range of parameters $\omega, b$ and $\gamma$.  A
typical solution is shown in Fig. \ref{int_num}. As one sees from the
Figure, for $b \ne0$ the asymptotic behavior of the interface exhibits
decaying oscillations.

Let us now consider the perturbed solution
\begin{equation}
{A \choose  B} =  {A_0 \choose  B_0} + { \tilde A (x) \choose  \tilde B(x)}   
\exp [\lambda(k) t
+ i k y].
\label{eq15}
\end{equation}
where $k$ is the transverse modulation wavenumber and $\lambda(k)$ 
is the corresponding growth rate. 
For $\tilde A , \tilde B$ we obtain linear equations
\begin{equation}
\hat L {\tilde A \choose  \tilde B}  =
(\lambda(k) + k^2) {\tilde A \choose  \tilde B} + b  k^2 {\tilde B
\choose -\tilde A}
\label{eq16}
\end{equation}
where the matrix $\hat L$ is of the form 
\begin{eqnarray}
\hat L =   \left (
\begin{array} {rl}
s -1 -3  A_0^2- B_0^2 +  \partial_x^2,   & -2  \omega
-2 A_0  B_0-b  \partial_x^2 \\
-2  A_0   B_0+b  \partial_x^2  ,  & -  s - 1-  A_0 ^2
- 3 B_0 ^2+ \partial_x^2
\end{array}
\right)
\label{L}
\end{eqnarray}

\begin{figure}
\centerline{\epsfxsize= 8.2cm\epsfbox{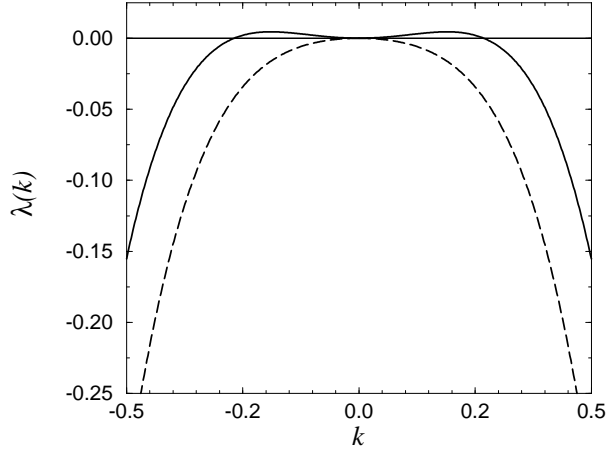}}
\caption{ Spectrum of eigenvalues $\lambda(k)$ for 
$\omega=1.3, b=4$ and $\gamma=2$ (solid line) and 
$\omega=1, b=4$ and $\gamma=2$ (dashed line)
obtained from numerical solution of 
Eqs. \protect{\ref{eq16}}.
}
\label{int_spec}
\end{figure}

In order to determine the spectrum of eigenvalues $\lambda(k) $ we  have
solved  Eq. (\ref{eq16}) along with stationary Eqs. (\ref{int_s})
numerically using numerical matching-shooting technique.  We have found
that for small $\omega$ the interface is stable with respect to
transverse undulations, i.e. $\lambda(k) \le 0$. However, for  the
values of $\omega$ above certain critical value $\omega_c(\gamma,b)$ the
interface exhibits transverse instability: $\lambda(k) > 0$ in the band
of wavenumbers  $|k| <k_c$, see Fig. \ref{int_spec}.  

\begin{figure}
\centerline{\epsfxsize=6.cm \epsfbox{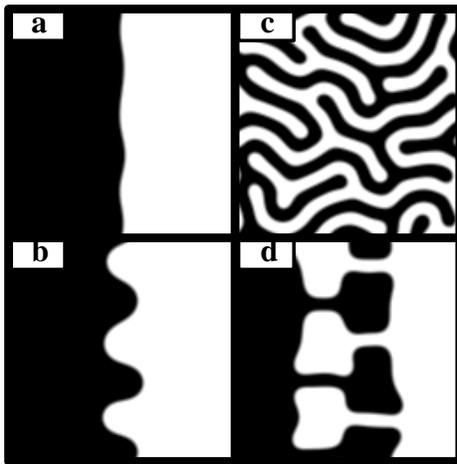}}
\caption{ (a-d) Interface instability  and labyrinth formation,
Eq. (\protect \ref{eq1}), $\omega=2, b=4, \gamma=2.9$, domain size
$100\times 100$ units, the snapshot are taken at times $t=1000, 1600, 2300, 
4640$.
 }
\label{interf_fig}
\end{figure}

This instability is confirmed by direct numerical simulations of
Eq.(\ref{eq1}). An example of the evolution of slightly perturbed
interface is shown in Fig. \ref{interf_fig}. Small perturbations grow to
form a ``decorated'' interface.  With time these decorations evolve
slowly and eventually form a labyrinthine pattern.

The neutral curve for this instability can
be determined as follows. Numerical analysis shows that at the threshold
the most unstable wavenumber is $k=0$ and we can expect that for $k \to 0 $
$\lambda  \sim k^2$, see Fig. \ref{int_spec}. 
Expanding Eqs.(\ref{eq16}) in power series
of $k^2$: 
\begin{equation} 
{\tilde A \choose  \tilde B} = {A^{(0)} \choose  B^{(0)}} +
 k^2 {A^{(1)} \choose  B^{(1)}} +...,
\end{equation} 
 in the zeroth order in $k$ we obtain
$\hat L (A^{(0)}, B^{(0})=0$.  The corresponding solution is the translation
mode $A^{(0)}=\partial_x  A_0(x), B^{(0)}= \partial_x  B_0(x)$.
In the first order in $k^2$ we arrive at the linear inhomogeneous problem
\begin{equation}
\hat L { A^{(1)} \choose   B ^{(1)}}  =
(\lambda(k) + k^2) { A^{(0)} \choose   B^{(0)}} + b  k^2 { B^{(0)}
\choose - A^{(0)}}
\label{eq18}
\end{equation}
A bounded solution to Eq. (\ref{eq18}) exists if the r.h.s. is orthogonal to
the localized mode of the adjoint operator $A^+, B^+$. The adjoint operator 
$\hat L^+$  is of the form
\begin{eqnarray}
\hat L^+ =   \left (
\begin{array} {rl}
s -1 -3  A_0^2- B_0^2 +  \partial_x^2,   &  -2  A_0   B_0+b  \partial_x^2 \\
-2  \omega
-2 A_0  B_0-b  \partial_x^2,  
& -  s - 1-  A_0 ^2
- 3 B_0 ^2+ \partial_x^2
\end{array}
\right)
\label{LA}
\end{eqnarray}
Since the operator $\hat L^+$ is not self-adjoint, 
the adjoint mode does not coincide with the translation mode, 
and, therefore, must be obtained numerically.

\begin{figure}
\centerline{\epsfxsize= 8.2cm\epsfbox{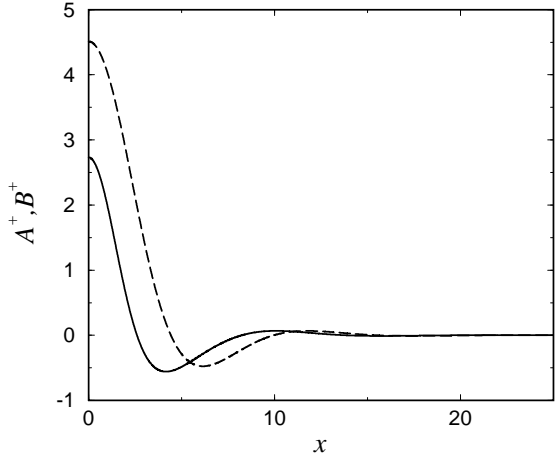}}
\caption{ Adjoint eigenfunction
obtained numerically for $\omega=1.3, b=4, \gamma=2$. Solid line corresponds
to $A^+$ and dashed line to $B^+$, respectively.}
\label{adj}
\end{figure}

The orthogonality
condition fixes the relation between $\lambda $ and $k$:
\begin{equation}
\lambda = -\sigma  k^2,
\sigma  = 1+ b \frac{\int_{-\infty }^\infty (A^{(0)} B^+ - B^{(0)} A^+) dx }
{ \int_{-\infty }^\infty (A^{(0)} A^+ + B^{(0)} B^+) dx }
\label{coef3}
\end{equation}
where $\sigma$ is "surface tension" of the interface. 
The instability, corresponding to the negative surface tension of the
interface,  onsets at $\sigma = 0$.  Although the interface solution
$A(x)$ is not localized (see Fig. \ref{int_num}), the integrals in Eq.
(\ref{coef3}) converge because the adjoint functions $A^+,B^+$ are
localized.  The numerically-obtained adjoint eigen-mode is shown in Fig.
\ref{adj}.  The neutral curve is shown in Fig. \ref{phase}. Direct 
numerical simulations of Eq.(\ref{eq1}) show that this
instability leads to  so called ``decorated'' interfaces (see Fig.
\ref{interf_fig},a), however   At later stages the undulations grow and
finally form labyrinthine patterns (Fig. \ref{interf_fig}, b-d).

The emergence of the labyrinthine patterns as a result of interface
instability contradicts experiments \cite{swin3,inter} where stable decorated 
interfaces are typically 
observed. Thus, the model Eq. (\ref{eq1}) does not capture saturation of the 
interface instability and requires modification. We also checked that 
the coupling to the density field introduced in Ref. \cite{at} also does not 
provide desired saturation. 

Let us now discuss possible mechanisms of saturation of the transversal
instability of the interface, which is not captured by Eq. (\ref{eq1}).
The reason for proliferation of the labyrinths is that 
local stabilization mechanisms cannot saturate the ``negative surface 
tension instability" of the interface. Indeed, the behavior of 
small perturbation  $\zeta$ of almost flat interface is described by the 
following linear equation: $\zeta _t = - \sigma  \Delta \zeta $, 
where $\sigma$ is the surface tension.  The linear 
growth must be counter-balanced by nonlinear terms. Due to translation 
invariance  the local nonlinearity can be function of gradients of 
$\zeta$ only. In the lowest order the first term is $|\nabla \zeta|^3$,
since the quadratic term does not saturate monotonic instability. 
Since for the modulation wavenumber $k\to 0$ one has 
$|\nabla \zeta|^3\sim k^2 |\zeta|^3$ and $\Delta \zeta \sim k^2  \zeta$,
the linear instability always overcomes local nonlinearity at small wavenumbers.
As a result small perturbations grow and saturation happens due to non-local
interactions as in a labyrinthine pattern which 
is stabilized by non-local repulsion of the fronts (see  \cite{goldstein}). 

As we see, in our model the interface instability does not result in stable 
decorated interfaces. One of the possibility for the saturation of 
this instability could be local  convection induced by the interface. 
Indeed, due to shear motion of grains at the interface (flat regions 
moves in anti-phase) local convective flow can be created. The scale 
of this flow can be large then the interface thickness. 
One of the possible manifestation  of the convection 
beside grain transport is
reduction of the effective "viscosity" of flat layer, or increase in 
the sensitivity to external vibrations\cite{bizon2}. As we can see from Fig. \ref{phase} 
increase in vibration magnitude $\gamma$ suppresses the interface instability. 
As the instability develops, the length of the interface grows, 
which increases the convection and finally suppresses the instability. 
This feedback mechanism results in saturation of the interface instability 
and creates stable decorations. 

Although the details of this coupling are not known yet, it can be implemented 
within the order-parameter approach in a phenomenological way. 
In Eq. \ref{eq1} the forcing $\gamma \psi^*$ 
must be replaced by $(\gamma_0+w) \psi^*$, where $w$ includes the
effect of local convective flow and $\gamma_0 = const$. 
We take the simplest possible form of the coupling to the convective flow: 
\begin{equation} 
w = \frac{\epsilon  }{S} \int d r^\prime  \exp \left(-
\frac{|r-r^\prime |^2 } {r_0^2} \right)  |\nabla \psi(r^\prime)|^2
\label{w} 
\end{equation} 
where $r_0$ characterizes the typical scale of the convective flow
and $\epsilon$  is the amplitude of the coupling, and 
$S$ is the area of integration domain. Since 
$ |\nabla \psi(r^\prime)|$ is nonzero only at
the interface, this integral is proportional 
to the total length of the interface. With the  growth of the 
interface length  the value of ``effective forcing" $\gamma= \gamma_0 + w$ 
grows as well, and eventually leaves the instability domain. 

\begin{figure}
\centerline{\epsfxsize= 8.2cm\epsfbox{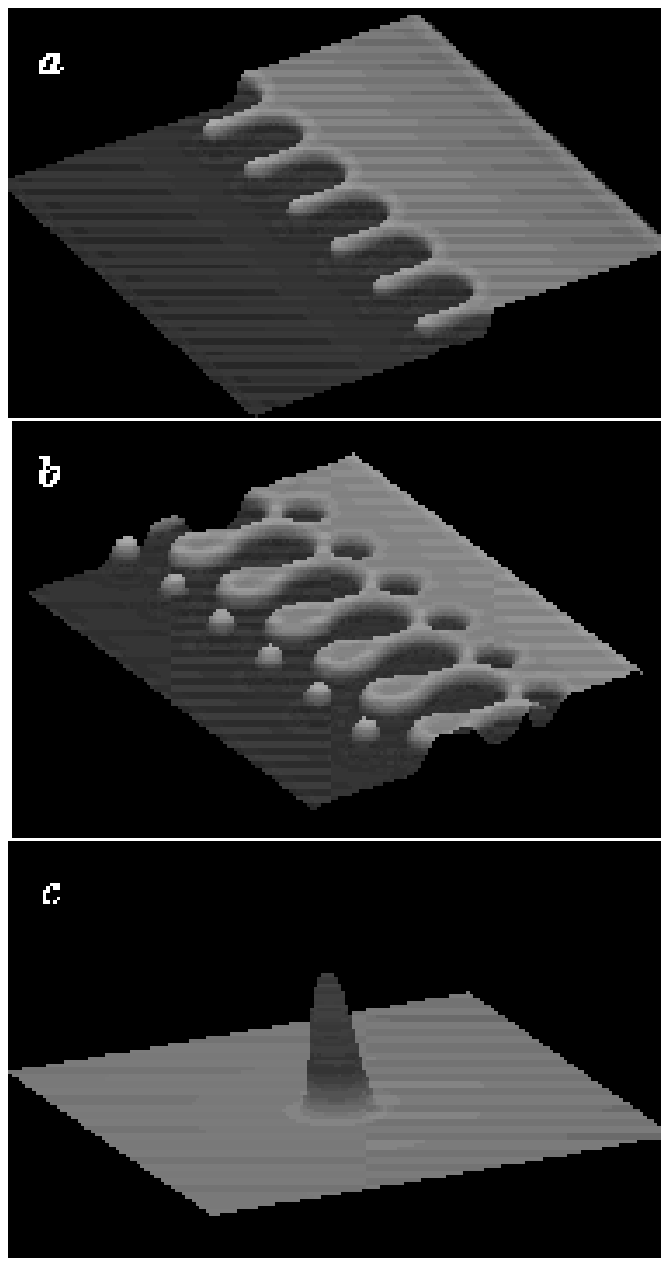}}
\caption{ Computer simulations of modified Eq. (\protect\ref{eq1}). 
Surface plots show $Re \Psi$ for different stationary solutions:
(a) the saturated interface, $\omega=1.2,\gamma=1.85, b=4$,
$\epsilon=0.002$, $r_0=40$, $L=160$, (b) saturated interface with bound
{\it super-oscillons},  $\omega=1.2,\gamma=1.75, b=4$, 
$\epsilon=0.002$, $r_0=40$, $L=160$. 
(c) single {\it super-oscillon} at $\omega=1.2,\gamma=2.2, b=4$, 
$\epsilon=0.002$, $r_0=40$, $L=100$ .}
\label{supos}
\end{figure}

We performed numerical simulations with Eq. (\ref{eq1}) modified in 
this fashion. 
The convolution integral (\ref{w}) was calculated using the fast 
Fourier transform. The results are presented in Fig. 
\ref{supos}. Remarkably, for certain range of parameters 
the instability indeed saturates due to nonlocal term Eq. (\ref{w}), 
and we find interfaces with stable decorations, see Fig. \ref{supos},a. 

\section{Localized Solutions} 
\begin{figure}
\centerline{\epsfxsize= 8.2cm\epsfbox{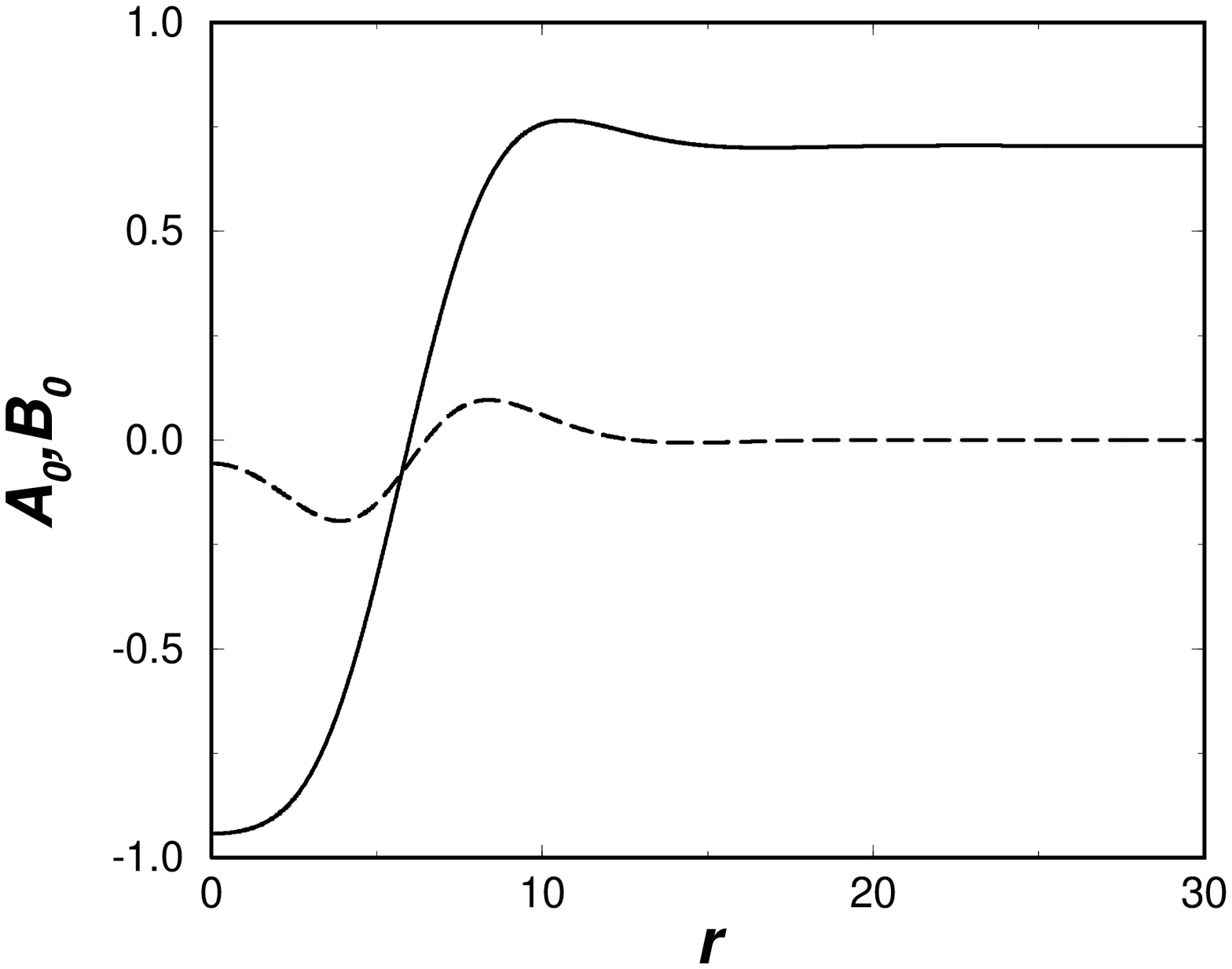}}
\centerline{\epsfxsize= 8.2cm\epsfbox{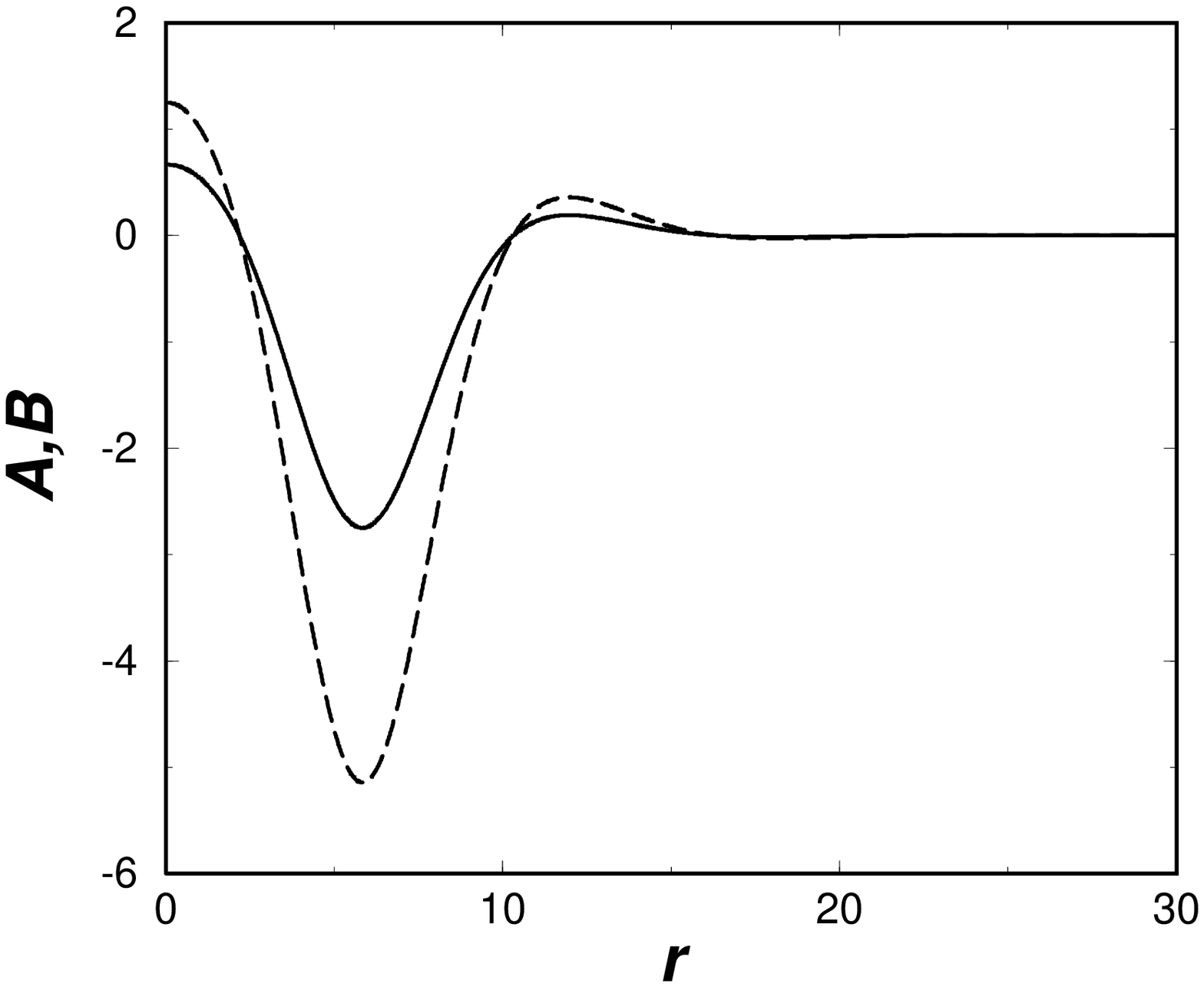}}
\caption{(a) Axi-symmetric 
localized solution (super-oscillon) obtained for 
$b=4,\omega=1$ and $\gamma=1.8$ for Eq. 
(\protect\ref{eq1}) at  $b=4$, $A$-solid line, $B$-
dashed line. (b) Localized eigenfunction corresponding 
to the maximum eigenvalue $\lambda =-0.0009$. 
} 
\label{sup_osc} 
\end{figure}

In addition to the interface solution which exists for $s>1$, 
Eq. (\ref{eq1}) for even higher values of $s$  possesses a variety of
other nontrivial two-dimensional solutions including localized
"super-oscillons". The typical solution and its corresponding localized
linear mode are shown in Fig. \ref{sup_osc}. 

Although super-oscillon appears similar to the ``conventional oscillon"
discovered in Ref. \cite{swin3}, there is a significant difference. 
Asymptotic behavior of the super-oscillon is 
$\Psi \to A_0\ne 0$ for $r \to \infty$ 
whereas the regular oscillon has  $\Psi\to 0$. As a result 
super-oscillons always have the same phase as their background. 
Oppositely-phased super-oscillons must be separated by the 
interface connecting two opposite phases of the period-doubled flat layer
whereas oppositely-phased oscillons can form stable bound states
(dipoles, \cite{swin3}). 

Super-oscillons exist in a relatively narrow domain of parameters of 
Eq. (\ref{eq1}). Decrease of $\gamma$ leads to destabilization of 
the localized eigen-mode leading to subsequent
nucleation of new "super-oscillons"
on the periphery of the unstable super-oscillon, with the increase of 
$\gamma$ super-oscillons disappear in a saddle-node bifurcation. 

Interestingly, for certain values of $\gamma_0$ we also find an interesting
structure consisting of a decorated interface and a chain of super-oscillons 
bound to the decorations, see Fig. \ref{supos},b. These objects
can exist completely independently of the interface, Fig. \ref{supos} c,
as one may expect for Eq. (\ref{eq1}). Thus, we conclude that 
extended Eq. (\ref{eq1}) qualitatively describes  observed phenomenology.  

\section{Phase diagram} 
\begin{figure}
\centerline{\epsfxsize= 8.2cm\epsfbox{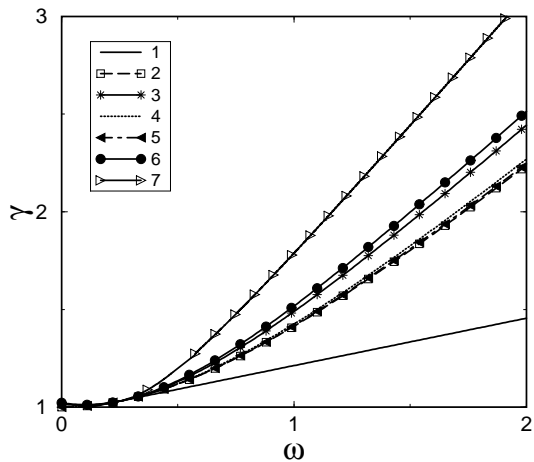}}
\caption{Phase diagram for Eq. (\protect\ref{eq1}) at  $b=4$. Line 1,
$\gamma^2=(\omega+b)^2/(1+b^2)$; line 2, $s^2\equiv
\gamma^2-\omega^2=1$; line 3, $s=s_c$; line 4, $s=s_c-\epsilon_R$; line
5, $s=s_c-\epsilon_B$;  line 6, $s=s_c-\epsilon_A$;  line 7, surface tension
$\sigma=0$. Below line 1 there are no patterns, between lines 1 and 4
stripes are stable, between lines 5 and 6 hexagons are stable, above
line 2, non-trivial flat states exist, and above line 3 they are stable.
Interfaces are unstable below line 7. 
Hexagons and stripes coexist between lines $4$ and $5$}
\label{phase}
\end{figure}

The results of linear stability analysis can be summarized in the phase
diagram shown in Fig. \ref{phase}. Remarkably,  the structure of the
phase diagram Fig.\ref{phase} is qualitatively similar to that of the
experiments for high frequencies of vibration, see Fig. \ref{phd} and
Refs. \cite{swin1,swin2,swin3,inter,bizon1,blair}.  Increasing vibration
amplitude $\gamma$ leads to transition from a trivial state to stripes,
hexagons, decorated interfaces, and finally, to stable interfaces.
Transition from unstable to stable interfaces also occurs with
decreasing $\omega$ (increasing vibration frequency $f$), in agreement
with Ref.\cite{blair,umban}.  In experiments, at yet higher $\Gamma$,
quarter-harmonic patterns appear, however these patterns are not
described by our model.  Our analysis also predicts co-existence of
stripes and hexagons in a very narrow parameter range (see Sec. II).
Super-oscillons were found in a narrow domain close to line 7 of Fig.
\ref{phase}. 

Note that square patterns which are emerge at
lower frequencies, are not included in the phase diagram.  They are
obtained within the full model Eqs.(\ref{eq0a}),(\ref{eq0b}) with
additional density field $\rho$, see \cite{at}.  

\section{Swift-Hohenberg equation} 
Near  the line $s=1$  (see Fig.  \ref{phase}) Eq. (\ref{eq1}) can be
simplified.
%The  evolution of the
%interface can be investigated near the line $s=1$  (see Fig.  \ref{phase}).
In the vicinity of this line  $ A \sim (s-1)^{1/2} $ and $ B \sim (s-1)^{3/2}
\ll A$.
In the leading order we can obtain
from Eq. (\ref{eq2b})  $B=b\nabla^2A/2$, and Eq. (\ref{eq2a})
yields \cite{coullet}:
\begin{equation}
\partial_t A =(s -1) A - A^3  + (1-\omega b) \nabla^2  A-\frac{ b^2}{2}
 \ \nabla^4 A.
\label{sh}
\end{equation}
Rescaling the variables $t \to (s -1 ) t,$
$A \to (s-1)^{-1/2}A,$ $x \to (2(s-1)/b^2)^{1/4}$, we  arrive 
to the Swift-Hohenberg equation (SHE)
\begin{equation}
\partial_t A = A - A^3  - \delta  \nabla^2  A-
 \ \nabla^4 A
\label{sh1}
\end{equation}
where
\begin{equation}
\delta =\frac{  \omega b - 1 }{b}  \sqrt {\frac{  2 }{s-1}}.
\label{delta}
\end{equation}
The description by SHE is valid if $\delta \sim O(1)$ which implies additional
condition for the validity of this approach $\omega b -1 \ll1 $
at $s \to 1$. 

Although  this equation is simpler than Eq. (\ref{eq1}), 
it captures many essential features of the original system dynamics,
including existence and stability of stripes and hexagons in
different parameter regions (see \cite{dewel,crawford1}),
existence of the interface solutions, interface instability, 
super-oscillons  and emergence of labyrinthine patterns 
\cite{ampt}. Indeed, simple analysis shows that
the growth rate of the instability of the uniform state $A=1$ as a
function of the perturbation wavenumber is
determined by the formula $\lambda (k) = -2 + \delta k^2 -k^4$, so it
becomes unstable at  $\delta>\delta_c=2\sqrt{2}$ at critical wavenumber
$k_c=\sqrt{2}$. As in the original model, near the threshold of this
instability, subcritical hexagonal patterns are preferred.
Interface stability can also be analyzed  more simply as the linearized
operator corresponding to the model (\ref{sh1}) is self-adjoint. The
threshold value of $\delta$ is obtained from the following
solvability condition
\begin{equation}
\int_{-\infty} ^ {\infty}  \left( \delta_{\mbox{th}} A_{0x}^2 - 2 A_{0xx}^2
\right)  d x =0
\end{equation}
which yields  $\delta_{\mbox{th}} =1.011$.
It yields the equation for the neutral curve of the interface instability
in the Swift-Hohenberg limit in the form: 
\begin{equation}
s=1+\frac{2 (\omega b -1 )^2}{\delta_{\mbox{th}} ^2 b^2}
\label{int_pos} 
\end{equation}

\begin{figure}
\centerline{\epsfxsize=6.cm \epsfbox{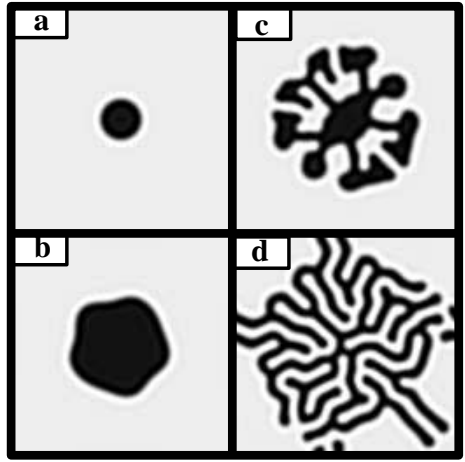}}
\caption{ 
(a-d) labyrinth formation from a circular spot, SHE, $\delta=1.4$,  domain size
$100 \times 100$ units, $t=200, 1300, 1700, 1900$.
 }
\label{interf_fig1}
\end{figure}

Fig. \ref{interf_fig1}, a-d  shows the development of the interface
instability within SHE with subsequent transition to labyrinthine
patterns, similar to the dynamics of the full parametric equation
(\ref{eq1}). Again, to saturate the instability, an additional
non-local mechanism is required.

As it was shown recently, Swift-Hohenberg equation also possesses
localized solutions (see \cite{ampt,staliunas}). These localized states
are analogous to the super-oscillons of Eq. (\ref{eq1}).

\section{Subharmonic forcing and interface motion}

Let us focus on the effect of additional subharmonic driving 
on the motion of the interface. 
In  the original model
Eq. (\ref{eq1}) the interface does not move  due to symmetry
between left and right halves of the interface.
However, if the plate oscillates with two frequencies, $f$ and
$f/2$, the symmetry between these two states connected by the interface, is
broken, and interface moves. The velocity of interface motion depends
on the relative phase of the sub-harmonic forcing with
respect to the forcing at $f $. The effect of a small external
subharmonic driving
applied on the back-ground of primary harmonic driving 
with the frequency $f$ can be described by the following equation
(cf. Eq. (\ref{eq1})) 
\begin{eqnarray}
\partial_t\psi&=&\gamma\psi^*-(1-i\omega)\psi+(1+ib)\nabla^2\psi
-|\psi|^2\psi+ q e^{i\Phi}
\label{eq_11}
\end{eqnarray}
where $q$ characterizes the amplitude of the sub-harmonic pumping, and
$\Phi$ determines its relative phase.

For small $q$, we look for moving interface solution in the form
$\psi= \psi_0(x-vt) + q \psi_1(x-vt) + ... $
and $v=O(q)$, or, alternatively, 
\begin{equation}
{A \choose B} = {A_0(x-vt)
 \choose B_0(x-vt)} +q {a(x-vt) \choose b(x-vt)} + O(q^2)
\end{equation}

\begin{figure}
\centerline{\epsfxsize=8.cm \epsfbox{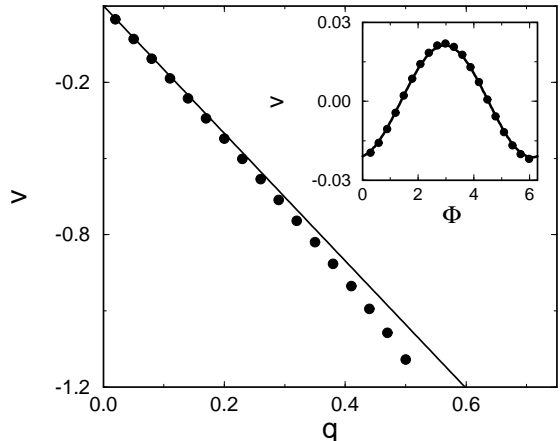}}
\caption{Interface velocity $v$ for $\omega=1, b=4, \gamma=2.5$
vs $q$ at $\Phi=0$.
Inset: $v$ vs  $\Phi$ at  $q=0.01$.
($\bullet$) - numerical results, (---) - analytical expression
(\protect \ref{speed}).
}
\label{interf_speed}
\end{figure}

Solvability condition (compare with Eq. (\ref{coef3})) fixes 
the interface velocity as a function of amplitude and the 
phase of external subharmonic driving
\begin{equation}
v=-q\frac{\cos\Phi \int{A^+dx} + \sin\Phi \int{B^+dx}}
{\int{(A^+\partial_xA_0+B^+\partial_xB_0)dx}}=q \alpha \sin(\Phi -\Phi_0) 
\label{speed}
\end{equation}
where $\alpha=const$ has the meaning of susceptibility 
of the interface for external forcing and $\Phi_0=const$ is some phase shift. 
The explicit answer is possible to obtain  for $b=0$  when $A^+=\partial_x
A_0$,
and $\Phi=0,\pi$, which yields
the interface velocity  $v=\mp \frac{3}{2}q A_0^{-2}=\mp
 \frac{3}{2}q (s-1)^{-1}$.
In general,  $A,B$, $A^+,B^+$, and hence $v$, can be
found numerically.  The
interface velocity as function of $q, \Phi$ is shown in Fig.
\ref{interf_speed}.

Thus, from this analysis we conclude that additional subharmonic driving 
results in controlled motion of the interface. The velocity of 
the interface depends on the amplitude of the driving, and 
the direction is determined by the relative phase $\Phi$.

\section{Experimental results} 

We performed experiments with a thin layer of granular material
subjected to periodic driving.  Our experimental setup is similar to
that of  \cite{inter}.  We used  $0.15$mm diameter bronze or copper
balls, and the layer thickness in our experiments was 10 particles. We
performed experiments in a rectangular cell of $4\times12$ cm$^2$.  We
varied the acceleration  $\Gamma$ and the frequency $f$ of the primary
driving signal as well as acceleration $\gamma$, frequency $f_1$ and the
phase $\Phi$ of the secondary (additional) driving signal. The interface
position and vertical accelerations were acquired  using a
high-speed video  camera
and accelerometers and further analyzed on a Pentium computer.  To
maintain and measure the proper acceleration at $f$ and $f/2$ we employ
the lock-in technique with our signal originating from an
accelerometer attached to the bottom of the cell.  This allows for the
simultaneous {\it real-time} feed back and control of the device.  To
reduce the interstitial gas effects\cite{pak} we reduced the pressure to
2 mTorr.  Our visual data is acquired using a high speed digital camera
(Kodak SR-1000c), in addition to high speed recording this also allows
for a synchronization between the patterns and the image acquisition.

\subsection{Phase Diagram} 
\begin{figure}
\centerline{\epsfxsize= 12cm\epsfbox{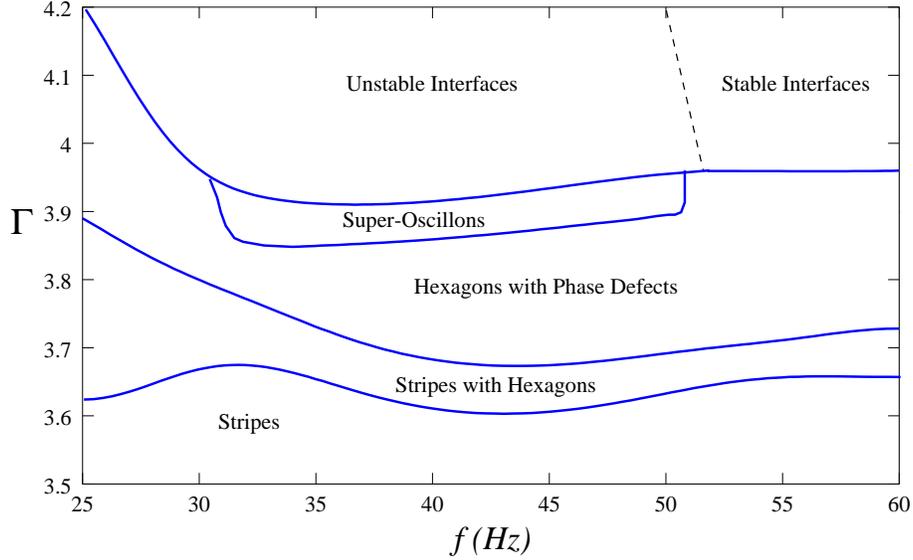}}
\caption{Phase diagram. Material is 150 $\mu m$ copper balls, 
layer thickness 10 particle diameters.}  
\label{phd} 
\end{figure}

\begin{figure}
\centerline{\epsfysize=7.2cm\epsfbox{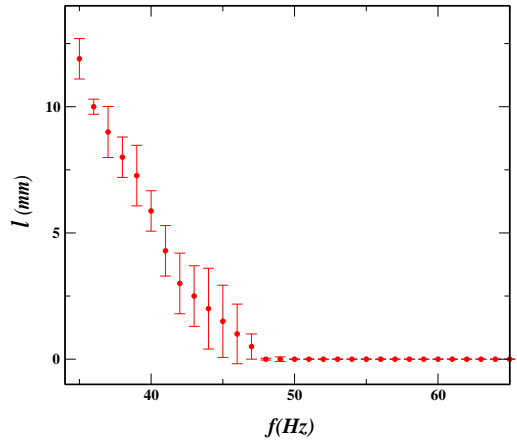}}
\caption{ The amplitude of periodic undulations 
$l$ as a function of driving frequency $f$.  
The acceleration amplitude $\Gamma=4.1$ }
\label{wid}
\end{figure}

The experimental phase diagram is shown in Fig. \ref{phd}. 
It is similar to that of Ref. \cite{swin2,swin3}, however the 
transition from hexagons to interfaces is elaborated in more detail. 
The dashed line
indicates the stability line for interfaces with respect to 
periodic undulations. To determine this 
stability limit,
we kept the value of acceleration  $\Gamma$ fixed and increased the
driving frequency $f$. For each value of the frequency we  extracted 
the interface width by averaging up to 10 images of the interface. 
In order to find that amplitude of the periodic undulations $l$  
we subtracted from the obtained value the 
thickness of the flat interface $l_0$.
The resultant amplitude or undulations $l$ 
as function of driving frequency $f$ is shown in Fig. \ref{wid}. 
As one sees from the Figure, the amplitude of undulations $l$
decreases gradually approaching the critical value of the 
frequency. However, a small hysteresis at the transition point 
cannot be excluded because of large error bars. 

\begin{figure}
\centerline{\epsfxsize= 8.2cm\epsfbox{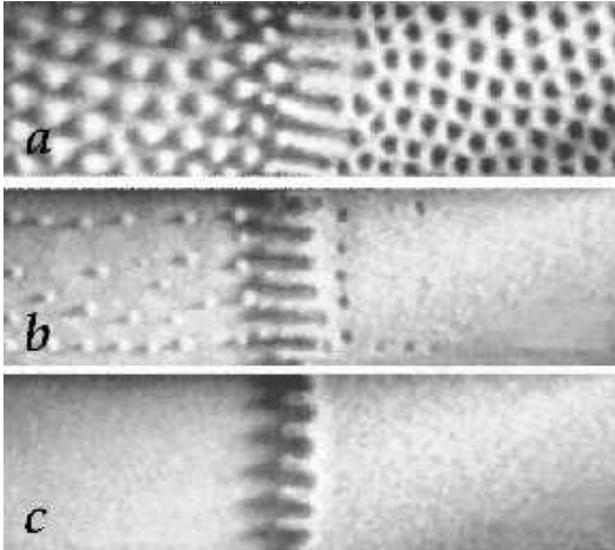}}
\caption{
Representative high-acceleration patterns in the 
rectangular $4\times12$ cm cell, 
driving frequency  $f=40$
Hz: (a) $\Gamma=3.75$ shows the onset of the
{\it interface} with up and down hexagons,
(b) $\Gamma=3.94$ {\it super-oscillons} are present in the
flat layer and are pinned to the front near the interface.
(c)  $\Gamma=4$, isolated decorated interface.  
}
\label{int_sup} 
\end{figure}

As it follows from the phase diagram,  
for small amplitude of the vertical acceleration $\Gamma<3.6$ 
stripes are the only stable
pattern (note that we focus on high-frequency patterns when squares do
not exist, compare \cite{swin2}).  At slightly higher $\Gamma=3.6..3.7$ 
stripes and hexagons coexist.  For higher $\Gamma=3.7..4.0$ the hexagons
becomes stable.  Due to the sub-harmonic character of motion both up and
down hexagons may co-exist separated by a line phase defect, Fig.
\ref{int_sup},a.  There exists a narrow band, from $\Gamma=3.85$ to
$\Gamma=4.0$, where localized states,  or {\it super-oscillons}, appear
both within the bulk of the material  and also pinned 
to the the front between the interface and the bulk Fig. \ref{int_sup},b. 
For even higher values of the acceleration $\Gamma>4.0$ the super-oscillons 
disappear (Fig.  \ref{int_sup},c) leading to isolated interfaces with
periodic decorations.  

\subsection{Controlled Motion of Interface} 

In the absence of additional
driving the interface drifts toward the middle of the cell
(see Fig. \ref{int_sup}). 
We attribute this  effect to the feedback between the
oscillating granular layer and  the plate vibrations due to
the finite ratio of the mass of the granular material to the
mass of the vibrating plate.
Even in the absence of subharmonic drive, the vibrating cell
can acquire subharmonic motion from the periodic impacts of the
granular layer on the bottom plate at half the driving frequency.
If the interface is located in the middle of the cell,
the masses of material on both sides of the interface are equal,
and due to the anti-phase character of the layer motion on both
sides, an additional subharmonic driving is not produced.
The displacement of the interface $X$ from the center of the cell leads 
to a mass difference $\Delta m$ on opposite sides of the interface 
which in turn causes
an additional subharmonic driving proportional to $\Delta m$.
In a rectangular cell $\Delta m \propto  X$.
Our experiments show that
the interface moves in such a way to decrease the subharmonic response,
and  the feedback provides an additional term $-X/\tau$ in the r.h.s. of
Eq. (\ref{speed}), yielding
\begin{equation}
\frac {d X}{ d t } = -  X/\tau +  q \alpha  \sin(\Phi - \Phi_0)
\label{eq2_1}
\end{equation}
The relaxation time constant $\tau$ depends on the mass ratio  (this
also holds for the circular cell if $X$ is small compared to the cell
radius).  Thus, in the absence of  an additional subharmonic drive
($q=0$), the interface will eventually divide the cell into two regions
of equal area (see Fig. \ref{int_sup}). 

\begin{figure}
\centerline{\epsfysize=3.5 in\epsfbox{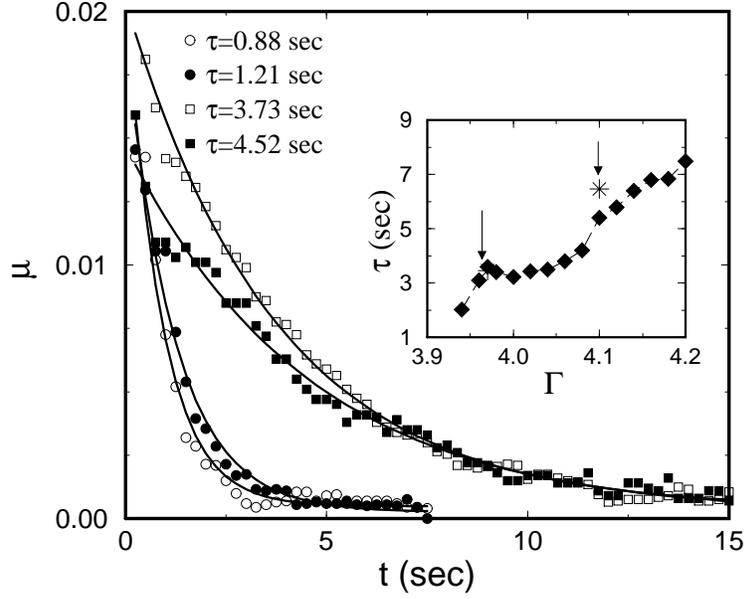}}
\caption{Amplitude of the  subharmonic acceleration
$\mu $ of  the cell averaged over 4 measurements vs
time for the interface propagating
to the center of the cell for $\Gamma = 3.97$ and
$f=40$ Hz.
Circles/squares correspond to the circular/rectangular
cells,  open/closed symbols correspond to
light/heavy cells, respectively. Heavy cells differ from
light ones by an additional weight of 250 g
attached to the moving shaft.
Solid lines show exponential fit
$\mu \sim  \exp (t/\tau ) + const $.
Inset: $\tau$ vs $\Gamma$ for light rectangular cell.}
\label{fig_rel} 
\end{figure}

In order to verify the prediction of Eq. (\ref{eq2_1}) we performed the
following experiment. We positioned the interface off center by applying
an additional subharmonic drive. Then we turned off the subharmonic
drive and immediately measured the amplitude $\mu$ of the plate
acceleration at the subharmonic frequency\cite{note}.  The results are
presented in Fig. \ref{fig_rel}.  The subharmonic acceleration of the
cell decreases exponentially as the interface propagates to the center
of the cell.  The measured relaxation time $\tau$ of the subharmonic
acceleration  increases with the mass ratio of the granular layer and of
the cell with all other parameters fixed.  The mass of the granular
layer was varied by using two different cell sizes: circular, diameter
$15.3$ cm, and rectangular, $4\times 12 $ cm, while keeping the
thickness of the layer unchanged.  For these  cells we found that the
relaxation time $\tau$ in the rectangular cell is about four times greater
than for the circular cell (see Fig. \ref{fig_rel}). This  is consistent
with the ratio of the total masses of granular material (52 grams in
rectangular cell and 198 grams in circular one).  In a separate
experiment the mass of the cell was changed by attaching an additional
weight of 250 grams  to the moving  shaft, which weights 2300 grams.
This led to an increase of the  corresponding relaxation times of 15-25
\%.  The relaxation time $\tau$  increases rapidly with $\Gamma$ (see
Fig. \ref{fig_rel} inset).

\begin{figure}
\centerline{\epsfysize=4 in\epsfbox{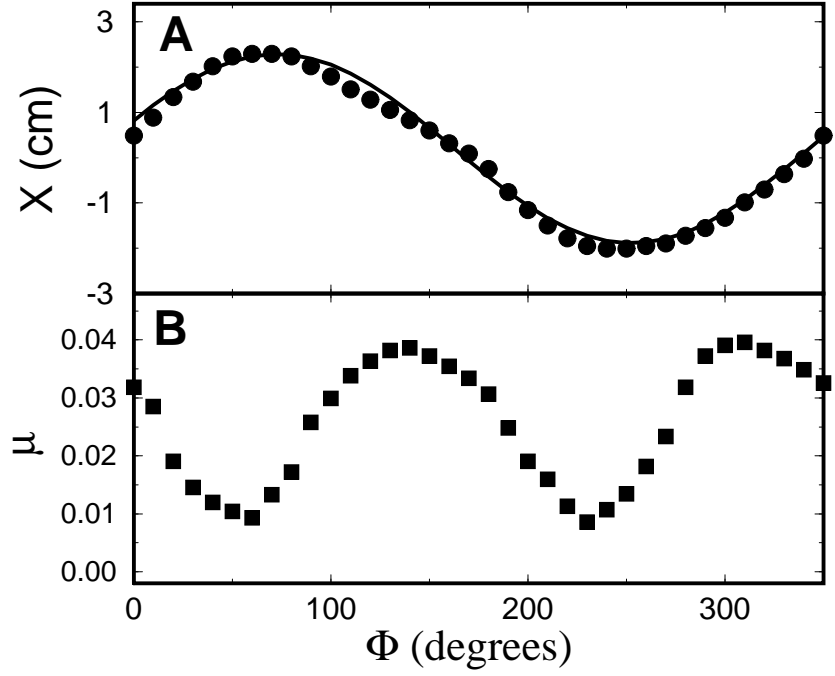}}
\caption{
(a)
Equilibrium position $X$ and (b) amplitude of measured
subharmonic acceleration $\mu$   as functions of phase $\Phi$.
Circular cell,
$\Gamma=4.1$, $f=40$ Hz, $q=1.25 \% $ of $\Gamma$.
}
\label{fig_accl} 
\end{figure}

When an additional subharmonic driving is applied, the interface is
displaced from the middle of the cell.  For small amplitude of the
subharmonic driving $\gamma$,  the stationary interface position is $X=
q \alpha \tau  \sin(\Phi - \Phi_0) $, since the restoring  force
balances the external driving force.  The equilibrium position $X$ as
function of $\Phi$ is shown in Fig. \ref{fig_accl}a.  The solid line
depicts the sinusoidal fit predicted by the theory.  Because of the
finite mass ratio effect, the amplitude of the measured plate
acceleration $\mu$ at frequency $f_1$ also shows periodic behavior with
$\Phi$, (see Fig. \ref{fig_accl}b), enabling us to  infer the interface
displacement from the acceleration measurements.  For even larger
amplitude of subharmonic driving (more than 4-5 \% of the primary
driving) extended patterns (hexagons) re-emerge throughout the cell.

\begin{figure}
\centerline{\epsfysize=4 in \epsfbox{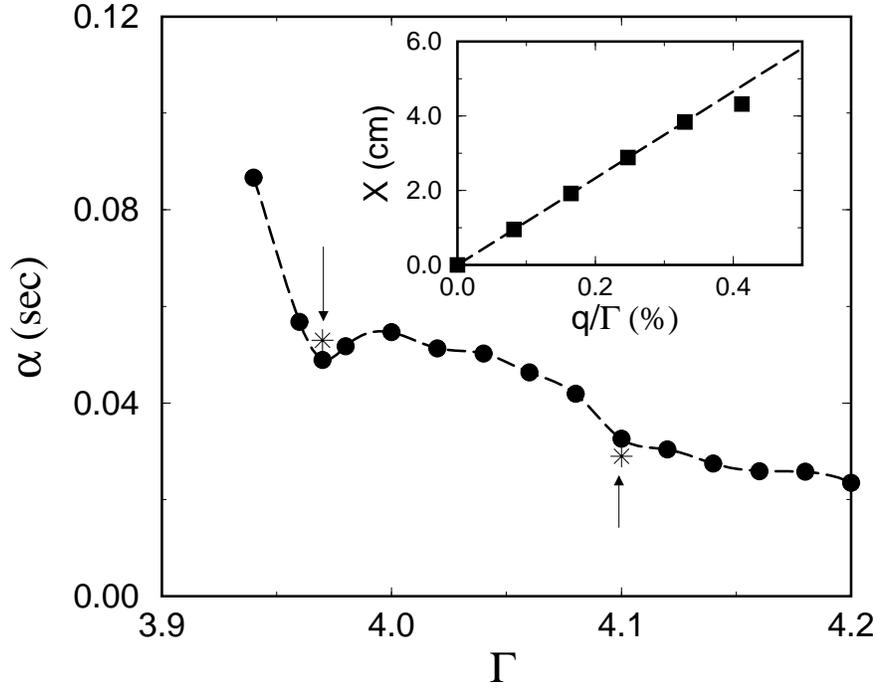}}
\caption{
Susceptibility $\alpha$ vs $\Gamma$
at $f=40$Hz,  rectangular cell.
Inset: Displacement $X$ as function of $q$ at $\Phi=260^0$.
}
\label{fig_dis}
\end{figure}

The velocity $V_0 =q \alpha $ which the interface would have in an
infinite system, can be found from the measurements of the relaxation
time $\tau$ and maximum displacement $X_m$ at a given amplitude of
subharmonic acceleration $\gamma$, $V_0=X_m/\tau$, see Eq.(\ref{eq2_1}).
We verified that in the rectangular cell the displacement $X$ depends
linearly on $\gamma$ almost up to values at which the interface
disappears at the short side wall of the cell (see Fig. \ref{fig_dis},
inset).  Figure \ref{fig_dis}  shows the susceptibility  $\alpha$ as a
function of the amplitude of the primary acceleration $\Gamma$. The
susceptibility decreases with $\Gamma$.  The cusp-like features in the
$\Gamma$-dependence of $\alpha$ (and $\tau$, see inset to Fig.
\ref{fig_rel}), are presumably related to the commensurability between
the lateral size of the cell and the wavelength of the interface
decorations.

\begin{figure}
\centerline{\epsfysize=4 in\epsfbox {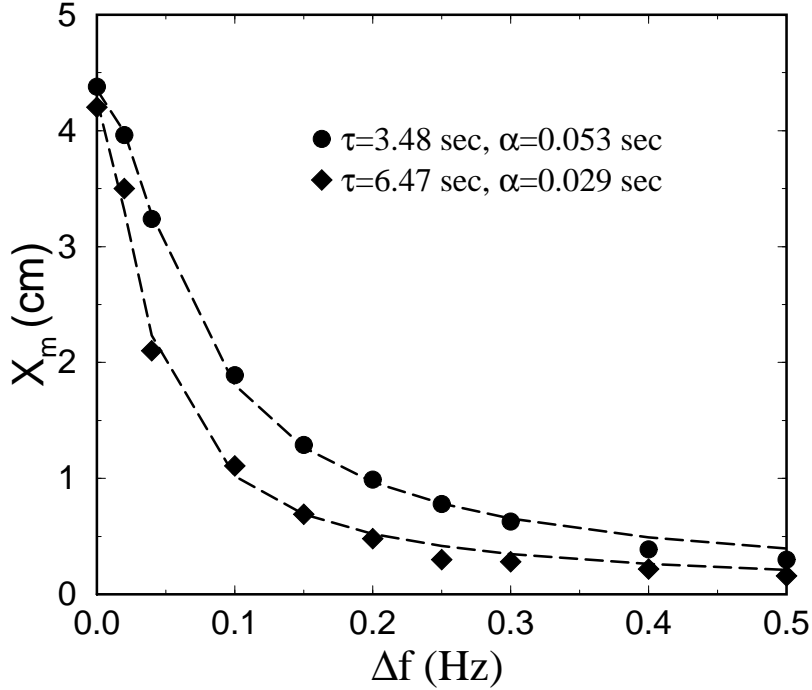}}
\caption{
Maximum displacement $X_m$ from center of rectangular
cell as function of
frequency difference $\Delta f = f_1 -f/2$
for $f=40$ Hz, and
for $\Gamma=3.97$ (circles) and $\Gamma=4.1$ (diamonds).
Dashed lines
are  fit to $X_m = V_0/\sqrt{\tau ^{-2} +
(2 \pi \Delta f)^2}$.
The values of $\alpha$ and $\tau$ 
obtained from the fit are also indicated in
Figs. \protect \ref{fig_rel} and \protect \ref{fig_dis} by arrows with stars.
}
\label{fig_de}
\end{figure}

We developed an alternative experimental technique which allowed us to
measure simultaneously the relaxation time $\tau$ and the ``asymptotic''
velocity $V_0$.  This was achieved by a small detuning $\Delta f$ of the
additional frequency $f_1$ from the exact subharmonic frequency $f/2$,
i.e. $\Delta f = f_1 -f/2 \ll f$.  It is equivalent to the linear
increase of phase shift $\Phi$ with the rate $2 \pi  \Delta f $. This
linear growth of the phase results in a periodic motion of the interface
with frequency $\Delta f $ and amplitude $X_m=V_0/\sqrt { \tau^{-2} + (2
\pi \Delta f)^2}$ (see Eq (\ref{eq2_1})).  The measurements of the
``response functions" $X_m(\Delta f)$ are presented in Fig.
\ref{fig_de}.  From the  dependence of $X_m$ on $\Delta f$  we can
extract parameters $V_0, \alpha$ and $\tau$ by a fit  to the theoretical
function.  The measurements are in a very good agreement with previous
independent measurements of relaxation time $\tau$ and susceptibility
$\alpha$.  For comparison with the previous results, we indicate the
values for $\tau$ and $\alpha$, obtained from the response function
measurements of Figs.  \ref{fig_rel} and \ref{fig_dis} (stars). The
measurements agree within  $5$ \%.

\section{conclusions} 
In this paper we  studied the dynamics of thin
vibrated granular layers in the framework of phenomenological model
based on the Ginzburg-Landau type equation for the order parameter
characterizing the amplitude of sub-harmonic sand vibrations. Although
rigorous derivation of this model is not possible up to date, our
approach gives a significant insight in the problem  in hand and results
in testable predictions.  Our model is complementary to more elaborate
fluid dynamics-like continuum descriptions or direct molecular dynamics
simulations and allows us to carry out analytical calculations of
various regimes and predict their stability. A challenging problem is to
establish a quantitative relation between our order parameter model and
other models and experiments.

We have shown in this paper that the generic parametric Ginzburg-Landau model
Eq.(\ref{eq1}) captures not only patterns of vibrated sand near the
primary bifurcation, but also large acceleration patterns such as
hexagons, interfaces, and super-oscillons.  The structure of the phase
diagram Fig.\ref{phase} for high frequencies and amplitudes of
vibrations is qualitatively similar to that of previous experiments
Refs.\cite{swin1,swin2,inter,blair}, as well as the experiments reported
here.  Increasing vibration amplitude leads to transition from a trivial
state to stripes, hexagons, decorated interfaces, and finally, to stable
interfaces.  Interface 
decorations were found to be a result of the transversal instability of
a flat interface which is analogous to the negative surface tension. 
We found that Eq. (\ref{eq1}) is not sufficient to describe
the saturation of interface instability. We propose a possible non-local
mechanism of saturation of this instability which takes into account the
dependence of the overall length of the interface and magnitude of
parametric forcing, which may occur due to a large scale convective flow
induced by the sand motion near the interface. 
We described analytically the motion of the interface under the
symmetry-breaking influence of the small subharmonic driving. 

In our experimental study we found that on a qualitative level 
the theoretical phase diagram is similar to the experimental one.  The
experiments also confirmed existence of super-oscillons and their bound
states with the interface as it was predicted by the theory.  
Further
experimental study is necessary to elucidate the specific mechanism for
saturation of the interface instability.

Stimulated by the theoretical prediction, we also performed experimental
studies of interface motion under additional subharmonic driving.  The
experiment confirmed  that  the direction and magnitude of the interface
displacement depend sensitively on the amplitude and the relative phase
of the subharmonic driving.  Moreover, we found  that period-doubling
motion of the flat layers produces subharmonic driving because of the
finite ratio of the mass of the granular layer and the cell. This in
turn leads to the restoring force driving the interface towards the
middle of the cell.

We thank L. Kadanoff, R.  Behringer,  H. Jaeger,  H. Swinney and P.
Umbanhowar for useful discussions.  This research was supported by the
US Department of Energy, grant \#  W-31-109-ENG-38, and by NSF, STCS
\#DMR91-20000. C.J. acknowledges support from joint ANL-UofC collaborative
grant.  L.T. thanks the Engineering Research Program of the Office of
Basic Energy Sciences at the U.S.  Department of Energy, grants
\#DE-FG03-96ER14592, DE-FG03-95ER14516 for support.

\end{document}